\documentclass[preprint,jcp]{revtex4}
\usepackage{graphicx}

\begin{document}

\title{Oxidation States of Graphene: Insights from Computational Spectroscopy}

\author{Wenhua Zhang$^{1,2,3}$}
\author{Vincenzo Carravetta$^3$}
\author{Zhenyu Li$^{1,2}$}
\thanks{E-mail:  zyli@ustc.edu.cn}
\author {Yi Luo$^{1,2}$}
\thanks{E-mail:  luo@kth.se}
\author {Jinlong Yang$^{1}$}

\affiliation{$^1$Hefei National Laboratory for Physical Sciences at
Microscale, University of Science and Technology of China, Hefei,
Anhui 230026, China}
\affiliation{$^2$Department of Theoretical Chemistry, School of
Biotechnology, Royal Institute of Technology, S-10691 Stockholm,
Sweden}
\affiliation{$^3$Institute of Chemical Physical Processes, CNR, via
Moruzzi 1, 56124 Pisa, Italy}

\begin{abstract}
When it is oxidized, graphite can be easily exfoliated forming graphene oxide (GO).
GO is a critical intermediate for massive production of graphene, and it is also
an important material with various application potentials. With many different
oxidation species randomly distributed on the basal plane, GO has a
complicated nonstoichiometric atomic structure that is still not well understood
in spite of of intensive studies involving many experimental techniques.
Controversies often exist in experimental data interpretation. We report
here a first principles study on binding energy of carbon 1\emph{s} orbital in GO.
The calculated results can be well used to interpret experimental X-ray
photoelectron spectroscopy (XPS) data and provide a unified spectral assignment. Based
on the first principles understanding of XPS, a GO structure model containing new
oxidation species epoxy pair and epoxy-hydroxy pair is proposed. Our results demonstrate
that first principles computational spectroscopy provides a powerful means to
investigate GO structure.
\end{abstract}

\maketitle

\section{Introduction}

Graphene is one of the most studied materials
in recent years owing to its peculiar physical properties and great potential
for various applications.
\cite{Geim:07:NMat, Castro-Neto:09, Geim:09} A big technological challenge in graphene
research is to massively produce high quality samples. Epitaxial growth
represents a suitable way for electronics application. \cite{Berger:06, Sutter:08, Kim:09:Nature, Li:09:Science}
However, for general purpose applications, solution based strategy is more attractive. One promising graphene
synthesis route is oxidizing and exfoliating graphite, then reducing the obtained
graphene oxide (GO) sheets. \cite{Ruoff:06:Nature, Car:07,
Tung:08:NNano, Chhowalla:08:Nanotech, GOreview} By this way, the pristine graphene
properties can be largely restored during the reduction.
At the same time, GO itself is an important material with various applications.
It is a good candidate for composite materials, \cite{Ramanathan:08} and its hydrate, GO paper,
shows excellent mechanic properties. \cite{Dikin:07} It is
possible to integrate GO into all-graphene electronic devices \cite{Wu:08:PRL}
and to obtain near-UV to blue photoluminescence from GO. \cite{Eda:09:AM} GO based
hydrogen storage materials are suggested to be very promising. \cite{Wang:09:IECR, Wang:09:ACSNano, Psofogiannakis:09}
By using GO support, Scheuermann et al. \cite{Scheuermann:09:JACS} also achieved
much higher Pd catalysis activity compared to conventional Pd/C system.

Because of its importance, it is highly desirable to
understand the atomic structure of GO and the oxidation processes to
prepare it. There is a long history of GO structure studies.
\cite{Hofmann:39, Ruess:46, Scholz:69, Nakajima:94, Lerf:98:JPCB,
Dekany:06:CM-XPS} In the most popular structure model, GO is
constituted of epoxy and hydroxy groups attached to the carbon layer, and
terminated with hydroxy and carboxyl groups. Many experimental
techniques, such as nuclear magnetic resonance (NMR),
\cite{Lerf:98:JPCB, Dekany:06:CM-XPS, Ruoff:08, Gao:09:GO} x-ray photoemission
spectroscopy (XPS), \cite{Dekany:06:CM-XPS, Lee:06:JACS,
Ruoff:08:ACSNano, Fan:08:AdvMater, Yang:09:Carbon} and vibrational spectroscopy,
\cite{Kudin:08, Hontoria:95:Carbon} have been employed to explore the structure information.
However, the interpretation of experimental data turns out to be difficult, since
GO is a nonstoichiometric compound with randomly distributed
oxidation groups.

XPS is one of the best local structure probes to identify oxidation
species on GO. Experimental XPS for heavily oxidized GO have mainly
three peaks, namely $P_1$, $P_2$, and $P_3$ from lower to higher
binding energies. It is commonly accepted that $P_1$ comes from
$sp^2$ carbon, and $P_2$ is contributed by carbon atoms connected to
epoxy (C-E) and hydroxy (C-OH) groups. However, definitive
assignment for $P_3$ is currently not available. Szabo et al.
\cite{Dekany:06:CM-XPS} assigned this peak to ketones (C=O), while
Jeong et al. \cite{Lee:06:JACS} attributed it to edge carboxyl
(COOH) groups. Considering its high intensity, peak $P_3$ can not be
solely contributed by edge groups. On the other hand, the
existence of interior ketone species has been excluded by other
experiments. For example, a recent two dimensional NMR experiment
suggested that carbonyl groups are spatially separated from the
majority \emph{sp$^2$}, C-OH, and C-E carbons. \cite{Ruoff:08} This
result implies that C=O groups mainly distribute at GO edges, as
also suggested by a recent O 1s X-ray absorption spectrum
measurement. \cite{Lee:08:Euro} Therefore, both experimental XPS assignments
are questionable, and new species may exist in GO \cite{zhenyu:09:cut} that
has a contribution to $P_3$.

Recently, new functional groups such as epoxy pair (EP) and
manganate ester (C-MnO$_4^-$) have been proposed in studies on
oxidation induced cut of graphitic materials. \cite{zhenyu:09:cut,
Kosynkin:09:Nature} When epoxy chain breaks underlying C-C bonds, \cite{Li:06:PRL} additional
epoxy groups can attach to it forming EP groups.
EP can then transit to carbonyl pair (CP) and finally break
the carbon network. However, there is a
significant dissociation barrier for the first EP to CP transition, \cite{zhenyu:09:cut}
which makes EP can in principle be widely existed in GO. Besides these two
groups, to reach a tentative assignment of NMR spectrum, a new structure motif of
lactol rings is suggested to be existed at GO edge sites.
\cite{Gao:09:GO} However, XPS signatures of these new spices are
still unclear.

Theoretical simulation can provide useful information to obtain
correct spectroscopic assignments and thus a more reliable GO structure model.
Unfortunately, previous theoretical studies on GO mainly focus on
energetics \cite{Paci:07:JPCC, Katsnelson:08, zhenyu:09:cut, Gao:09:JACS, Yan:09:PRL} instead of
properties that can be directly compared with experimental data. The complexity of the GO potential
energy surface limits the power of energetics study to some extent, especially when artificial
periodic boundary condition must be adopted. Kudin et al. \cite{Kudin:08} have
simulated the Raman spectra of GO, and proposed an alternating
single-double carbon bond model. Compared to Raman spectroscopy, XPS
is a more localized probe and thus a better choice to study local
structures.

In this study, we calculate core chemical shift
of carbon 1\emph{s} orbital for possible oxidation species in GO
using density functional theory, which can be compared to experimental XPS spectra directly
and leads to a unified reassignment. Based on such a computational spectroscopy
strategy, we propose a GO structure model with new species epoxy pair and hydroxy-epoxy pair.
The remainder of this article is organized as follows. In
section 2, we introduce the computational methods. XPS simulation results
and implications to GO structure are present in section 3. Finally, we conclude
in section 4.

\section{Computational details}

\textbf{XPS simulation with plane wave basis set.}
Spin polarized electronic structure calculations
are performed with the Vienna ab-initio simulation package
(VASP) \cite{vasp1} using Perdew, Burke and Ernzerhof (PBE)
\cite{pbe} exchange-correlation functional. The electron-ion
interaction is described with projector augmented wave (PAW)
method \cite{Kresse:99:PAW}, which permits to set the occupation of
core level as any number between 0 and 1. Different two dimensional sheet
and one dimensional nanoribbon models are adopted to simulate GO.
The binding energy of C 1$s$ orbital is calculated
as the energy difference between the ground state and the
core-excited state with one core electron removed. \cite{Kresse:04}
We have tested this computational approach by calculating the 1$s$
core orbital binding energies of C, N, and O atoms in alanine and
tryptophan, and the obtained core chemical shifts in different
chemical environments agree well with the corresponding experimental
values. \cite{EPAPS}

\textbf{XPS simulation with atomic basis set.} Cluster models are
also used to calculated the binding energy of C 1$s$ orbital. The
geometries of all clusters are optimized with DMol$^3$ \cite{Dmol-1,
Dmol-2} and the binding energy calculations are performed with
StoBe-deMon. \cite{Stobe} In binding energy calculations, gradient
corrected Becke (B88)\cite{B88} exchange functional and Perdew
(P86)\cite{PD86} correlation functional are adopted. Effective core
potential is used to describe the non excited atoms and for ionized
atoms, IGLO-III basis set\cite{basis} is used instead. To get the
binding energy of core electron, one C 1\emph{s} electron is removed
from the system. The core chemical shift is defined as the energy
difference between core electron removed system and the ground
state. Such a XPS simulation protocol has been widely used in many
studies, and it generally leads to a good agreement with experiment.
\cite{amino:09:zhang, C48N12-03, Brena-06} For different cluster
models, the core chemical shift is calibrated according to the
binding energy of C-H far way from the oxidation groups.

\section{Results and discussion}

The main task of this study is to calculate the core chemical shift
of C 1\emph{s} orbital for different oxidation species. For this
purpose, two first principles simulation protocols and many
different structure models are used. The absolute value of core
chemical shift can be affected by many factors, for examples, by
charging effect in experiment and by work function of the
theoretical GO model. We find that for a perfect graphene sheet, the
work function has a value of 4.7 eV, while for a hydrogen saturated
1D graphene nanoribbon, it becomes 3.7 eV. It indicates that the
structure model used for GO can have a significant effect on the
work function and thus on the absolute value of core chemical shift.
Therefore, we only focus on relative core chemical shift (R-CCS).
The average binding energy of epoxide group in each structure model
is set as a common reference to compare results from calculations
and from experiments.

\begin{figure}[!htb]
\includegraphics[width=7.0cm]{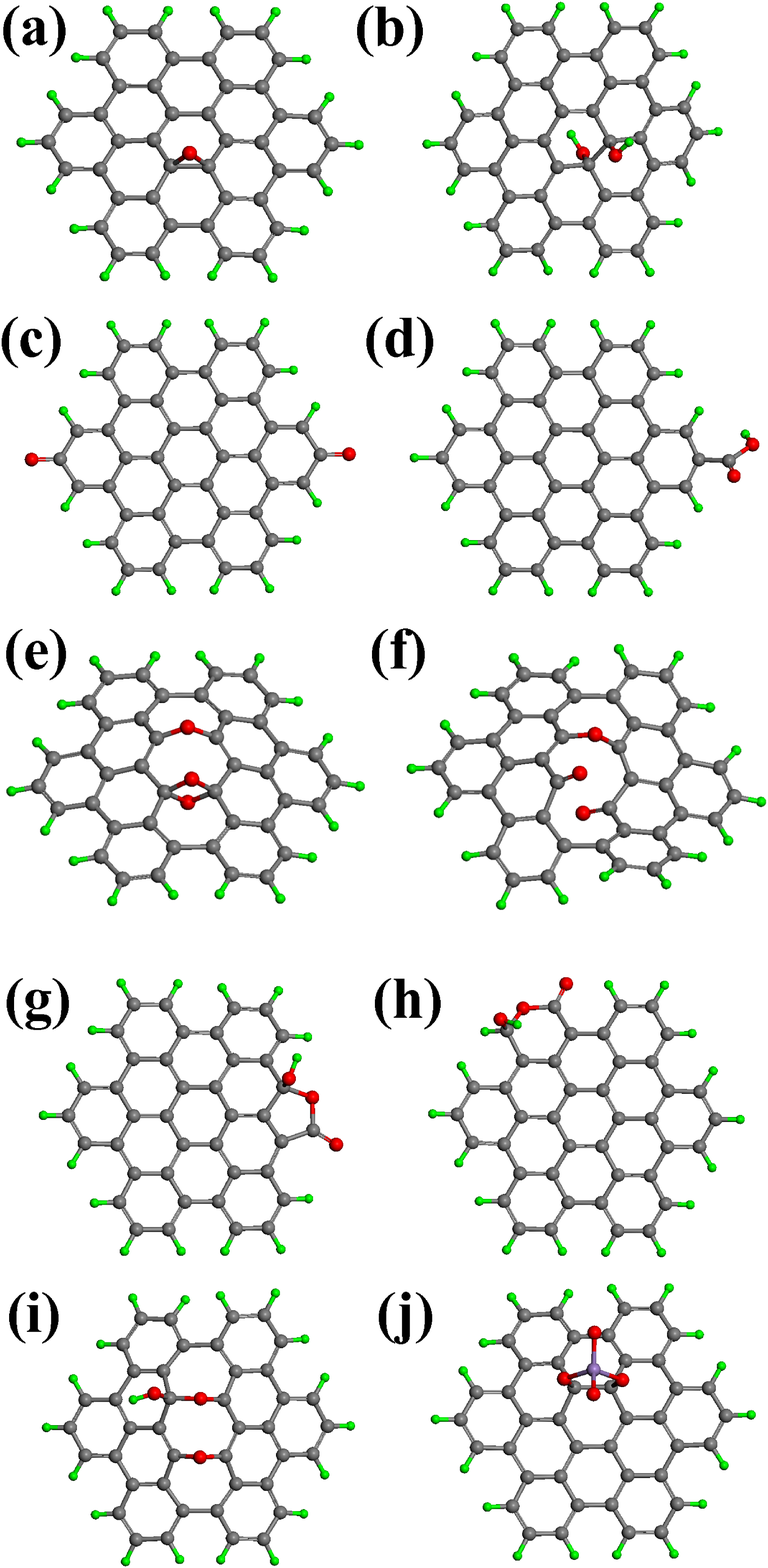}
\caption{Cluster models used to calculate C 1s binding energies.
(a) epoxy, (b) two OH groups,
(c) two edge C=O groups, (d) carboxyl group,
(e) line defect epoxy group (C-E$^\prime$) and epoxy pair (C-EP) group,
(f) C-E$^\prime$ and carbonyl pair (CP),
(g) five-ring lactol group, (h) six-ring lactol group,
(i) hydroxy-epoxy pair, and (j) MnO4- attached structure.}
\label{fig:0D}
\end{figure}

We first consider the well studied epoxy and hydroxyl groups, which contribute to XPS peak $P_2$.
In some experiments, $P_2$ is further deconvoluted into two sub-peaks, namely
$P_2^a$ and $P_2^b$. \cite{Lee:06:JACS, Ruoff:08:ACSNano} The former is about
0.8 eV lower in binding energy than the latter. These two peaks have been tentatively assigned
to C-OH and C-E, respectively. However, such an assumption needs more stringent test, for example,
by first-principles calculation on the binding energy of carbon 1$s$ orbital.

\begin{figure}[!htb]
\includegraphics[width=8.0cm]{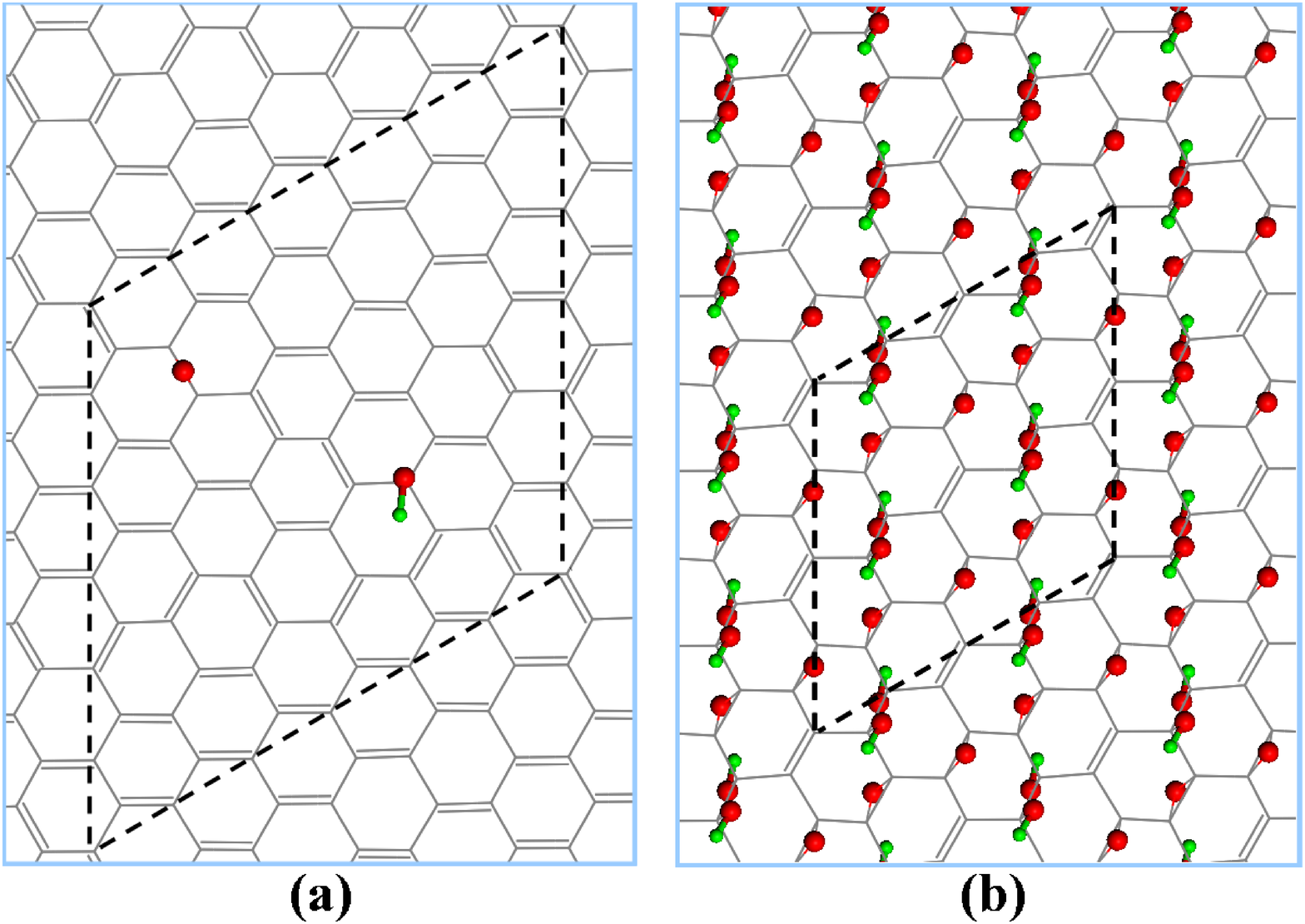}
\caption{(a) Graphene sheet with isolated epoxy and hydroxy groups.
Dashed lines mark a ($6\times 6$) unit cell. (b) A GO
model with epoxy and hydroxy groups. Dashed lines
indicate the ($4\times 4$) cell used in our calculations. Red and
green spheres represent oxygen and hydrogen atoms, respectively.}
\label{fig:2D}
\end{figure}

As shown in Figure \ref{fig:0D}a and \ref{fig:0D}b, we construct
two clusters, one with C-E and the other with C-OH. Both clusters have
an exclusive Clar's structure,
where all $\pi$-electrons form separated aromatic sextets. \cite{Clar:72}
Dangling bonds are saturated with hydrogen, and
the oxidation groups are put in the center of the cluster. The calculated R-CCS
of C-OH with respect to C-E is only 0.1 eV, indicating that C-E and C-OH are difficult to
be resolved in XPS. It is also interesting to note that the core binding energy of C-OH is
larger than that of C-E, contrary to the tentative assignment for $P_2^a$ and $P_2^b$.

For periodic systems, we put two isolated hydroxy and epoxy groups
in a big ($6\times 6$) supercell, as shown in Figure \ref{fig:2D}a.
R-CCS of \emph{$sp^2$} C with respect to C-E is -1.7 eV, while the
difference between the binding energies of C-E and C-OH is still within
0.1 eV, consistent with the atomic basis set calculation. To take
the coupling between epoxy and hydroxy groups into account, we also
consider a more realistic GO structure model \cite{Katsnelson:08}
(Figure \ref{fig:2D}b). In this case, the binding energy of C-OH is
larger than that of C-E by 0.3 eV, showing the important effect of chemical
environment. However, such a small difference is still not
distinguishable in XPS experiments. \cite{Fan:08:AdvMater,
Dekany:06:CM-XPS} Therefore, a new assignment for $P_2^a$ and
$P_2^b$ is required.

\begin{figure}[!h]
\begin{center}
\includegraphics[width=8.0cm]{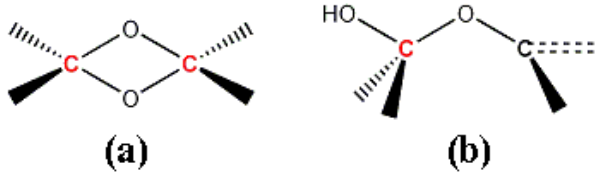}
\end{center}
\caption{New oxidation species. (a) epxoy pair, (b) hydroxy-epxoy pair. The
corresponding carbon atoms (red) are named as C-EP and E-C-OH, respectively.
Both C-EP and E-C-OH are monovalently connected to two oxygen atoms.} \label{fig:pair}
\end{figure}

\begin{figure}[!h]
\begin{center}
\includegraphics[width=8.0cm]{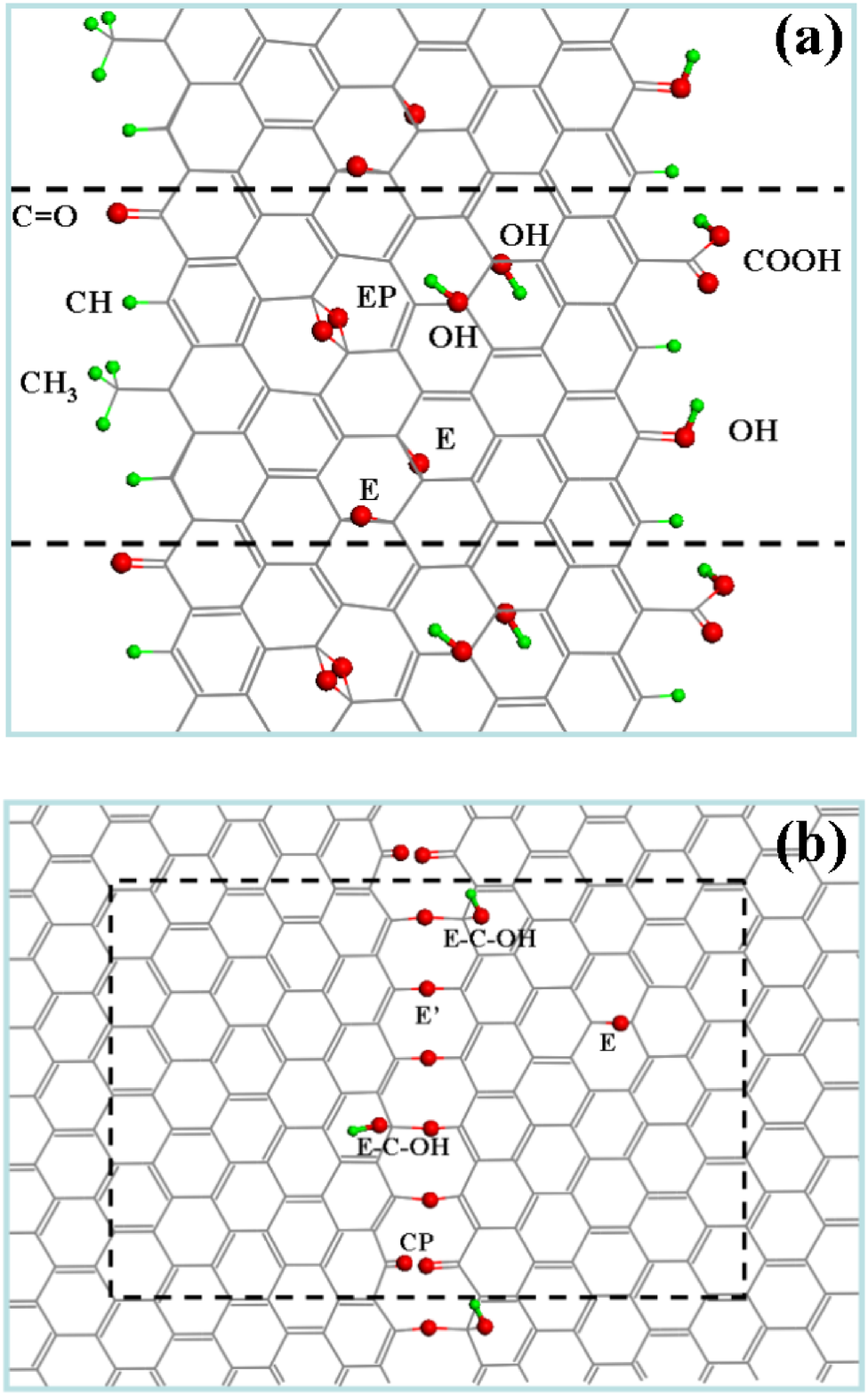}
\end{center}
\caption{(a) A GO nanoribbon model with zigzag edges. The part between
two dashed lines is a unit cell along the ribbon direction. (b) A GO structure
model based on epoxy line defect.} \label{fig:GO}
\end{figure}

To better understand the XPS peaks, more oxidation species beyond
C-OH and C-E should be considered. First, we take edge groups. It is known that
there are C=O, COOH, and hydroxy (C-OH-edge) groups existing at GO edge. In this study, we also consider
other edge groups including C-H, CH$_3$, CO$_3$, and lactol  (Figure \ref{fig:0D}g and \ref{fig:0D}h).
For new interior groups, some clues can be found from the oxidation induced cutting of
graphene. \cite{zhenyu:09:cut} In highly oxidized GO, some epoxy
groups may align in a line, which leads to the break of the underlying C-C bonds. \cite{Li:06:PRL}
Then, a new epoxy group can be attached to form a EP group (C-EP, Figure
\ref{fig:GO}a), \cite{zhenyu:09:cut} or a hydroxy can be added
to from a epoxy-hydroxy pair (E-C-OH, Figure \ref{fig:GO}b).
E-C-OH is actually also existed in the recently proposed edge lactol
groups, \cite{Gao:09:GO} and the latter can be considered as a
special case of the former.

Once the wide existence of aligned epoxy groups is recognized, the formation
of C-EP and E-C-OH becomes very natural. When aligned epoxy groups break the
underlying C-C bonds, the corresponding carbon atoms ($\alpha$ carbon) are
connected to two other carbon atoms ($\beta$ carbon) and the epoxy group, to which
the $\alpha$ carbon is bound with one valence electron. Therefore, there is
a $\pi$ electron left on the $\alpha$ carbon. If both the two $\beta$ carbon atoms are
oxidized to be $sp^3$ hybridized, then the $\alpha$ carbon becomes a radical. In
highly oxidized samples, the possibility that $\beta$ carbon atoms are saturated is quite high.
Therefore, the $\alpha$ carbon is very possible to
be attacked to form epoxy pair or epxoy-hydroxy pair groups in the oxidizing environment.

Together with C-E and C-OH, these new oxidation species have been put in
various of structure models. Individual groups have been put on clusters
(Figure \ref{fig:0D}). We also construct several GO nanoribbon models with many
oxidation species randomly distributed both on the plane and at the edges. One
of these ribbon models with zigzag edges is shown in Figure \ref{fig:GO}a.
It turns out that the calculated core chemical shifts are almost not
dependent on whether zigzag or armchair edges are adopted in the GO
model. Another kind of GO models we considered is epoxy line defect based, for which
different functional groups are put on the epoxy chain, as shown in Figure \ref{fig:GO}b.

With these GO models, we have calculated the core chemical shift for oxidation species
C-OH, C-E, C-EP, and E-C-OH on planes, and C-OH-edge, C=O, C-H, CH$_3$, COOH, COOO, and
lactol at edges. The calculated R-CCS values are collected in Table \ref{tab:BE},
together with experimental values reported in the literature. \cite{Dekany:06:CM-XPS,
Lee:06:JACS} Chemical shifts for all calculated oxidation species
are well within the energy window of the experimental XPS data.
Again, we find that epoxy and hydroxy groups in the interior of
ribbons are not distinguishable. The peak $P_2^a$ originally
assigned to C-OH by Jeong et al. \cite{Lee:06:JACS} has
contributions from edge attached C-OH-edge and C=O.

We notice that, in experimental spectra, $P_2^a$ have comparable intensity to $P_2^b$.
Therefore, there should be other contributions to $P_2^a$ besides the edge groups. For
epoxy chains with the underlying C-C bonds broken, the
corresponding carbon atoms (C-E$^\prime$) almost keep \emph{$sp^2$}
hybridization with a C-O bond length about 1.37 \AA, very close to
that in C-OH-edge (1.36 \AA). R-CCS of C-E$^\prime$ is also very
similar to that of C-OH-edge. Since it is distributed
in the interior part of GO, C-E$^\prime$ can make a strong peak in XPS.

Our calculations show both C-EP and E-C-OH contribute to the XPS peak $P_3$.
R-CCS of C-EP in an epoxy line defect (1.5 eV) is slightly higher than its
isolated counterparts (0.9 to 1.0 eV). E-C-OH has a larger
core level binding energy compared to C-EP. Edge group COOH has a similar R-CCS with
C-EP. Since the newly proposed five- or six-membered lactol ring contain
E-C-OH, its R-CCS also locates in the region of $P_3$, only slightly lower than
that of interior hydroxy-epoxy pair. However, as an edge groups
like COOH, it only has limited contribution to XPS. Main contribution to $P_3$
thus comes from C-EP and E-C-OH.

\begin{table*}[!h]
  \caption{R-CCS of carbon 1$s$ orbital for various oxidation species reported in
  experiment and calculated with atomic and plane wave basis sets in this work.
  The binding energy of C-E (287.2 eV) is set as a common reference point.
  The numbers in parentheses are the linewidths of experimental peaks.
  }
  \label{tab:CLS}
  \begin{tabular}{l | c  | c  | c  c  }
    \hline\hline
                           &Atomic           & Plane Wave          &EXP\cite{Dekany:06:CM-XPS} &EXP\cite{Lee:06:JACS} \\
    \hline
    C-H                    & -1.8            &-2.5 $\sim$ -1.9     &                      &                       \\
    C \emph{sp$^2$}        &-1.7             &-2.3 $\sim$ -1.6     &-3.1                  & -2.2 ($\pm 1.2$)      \\
    CH$_3$                 &                 &-1.8 $\sim$ -1.6     &                      &                       \\
    C-OH-edge              &                 &-0.9 $\sim$ -0.6     &                      & -0.8 ($\pm 1.0$)$^c$  \\
    C=O                    &-0.7             &-0.9 $\sim$ -0.1     &                      & -0.8 ($\pm 1.0$)$^c$  \\
    C-MnO$_4^-$            &-0.7             &                     &                      &                       \\
    C-E$^\prime$           &-0.7$\sim$-0.5   &-0.8 $\sim$ -0.3     &                      &                       \\
    C-OH                   &+0.1             &-0.1 $\sim$ +0.3     &0.0 ($\pm 1.2$)       &                       \\
    C-E                    &0.0              &-0.3 $\sim$ +0.3     &0.0 ($\pm 1.2$)       & 0.0 ($\pm 1.0$)       \\
    CP                     &0.0$\sim$+0.1    &-0.3 $\sim$ -0.2     &                      &                       \\
    C-EP                   &+1.1             &+0.9 $\sim$ +1.5     &+2.1 ($\pm 1.2$)$^a$  & +1.8 ($\pm 1.2$)$^d$  \\
    COO(H)                 &+1.3$\sim$+1.9   &+0.9 $\sim$ +1.4     &+2.1 ($\pm 1.2$)$^a$  & +1.8 ($\pm 1.2$)$^d$  \\
    E-C-OH                 &+1.7$\sim$+1.9   &+1.5 $\sim$ +1.9     &+2.1 ($\pm 1.2$)$^a$  & +1.8 ($\pm 1.2$)$^d$  \\
    COOO                   &                 &+3.0                 &+3.6 ($\pm 1.0$)$^b$  &                       \\
    \hline\hline
\end{tabular}\\
\noindent $^a$ assigned to carbonyl groups in Ref. \cite{Dekany:06:CM-XPS}. \\
\noindent $^b$ assigned to carboxyl groups in Ref. \cite{Dekany:06:CM-XPS}. \\
\noindent $^c$ assigned to hydroxy groups in Ref. \cite{Lee:06:JACS}. \\
\noindent $^d$ assigned to carboxyl groups in Ref. \cite{Lee:06:JACS}.
\label{tab:BE}
\end{table*}

EP can dissociate to form a carbonyl pair (CP, Figure \ref{fig:0D}f and \ref{fig:GO}b).
\cite{zhenyu:09:cut} R-CCS of an individual CP is close
to zero, significantly higher than edge C=O group, indicating its chemical environment
is strongly affected by neighboring C-E$^\prime$. If the whole line defect is broken,
CP becomes C=O edge groups with a negative R-CCS. Since an existing CP strongly lowers
the dissociation barrier of neighboring EPs, isolated CPs are not expected to be widely
existed.\cite{zhenyu:09:cut} However, an important fact disclosed here it that,
even if C=O groups exist in the interior of GO, its contribution to XPS
goes to $P_2$ instead of the originally supposed  $P_3$. \cite{Dekany:06:CM-XPS}

Another spices mentioned in the literature is C-MnO$_4^-$. In our
VASP calculations with plane wave basis set, for both a (6$\times$6)
unitcell and a cluster model, we fail to get a local minimum
structure with MnO$_4^-$ adsorbed on graphene, MnO$_4^-$ always
moves away during the optimization. This result suggests that
C-MnO$_4^-$ may be not stable or only stable in solution, and thus
does not exist during XPS measurements. An adsorbed structure can be
obtained using the atomic basis set with the DMol$^3$ package. With
this geometry, the calculated R-CCS is about -0.7 eV, far away from
peak $P_3$.

\begin{figure}[tbh]
\includegraphics[width=9cm]{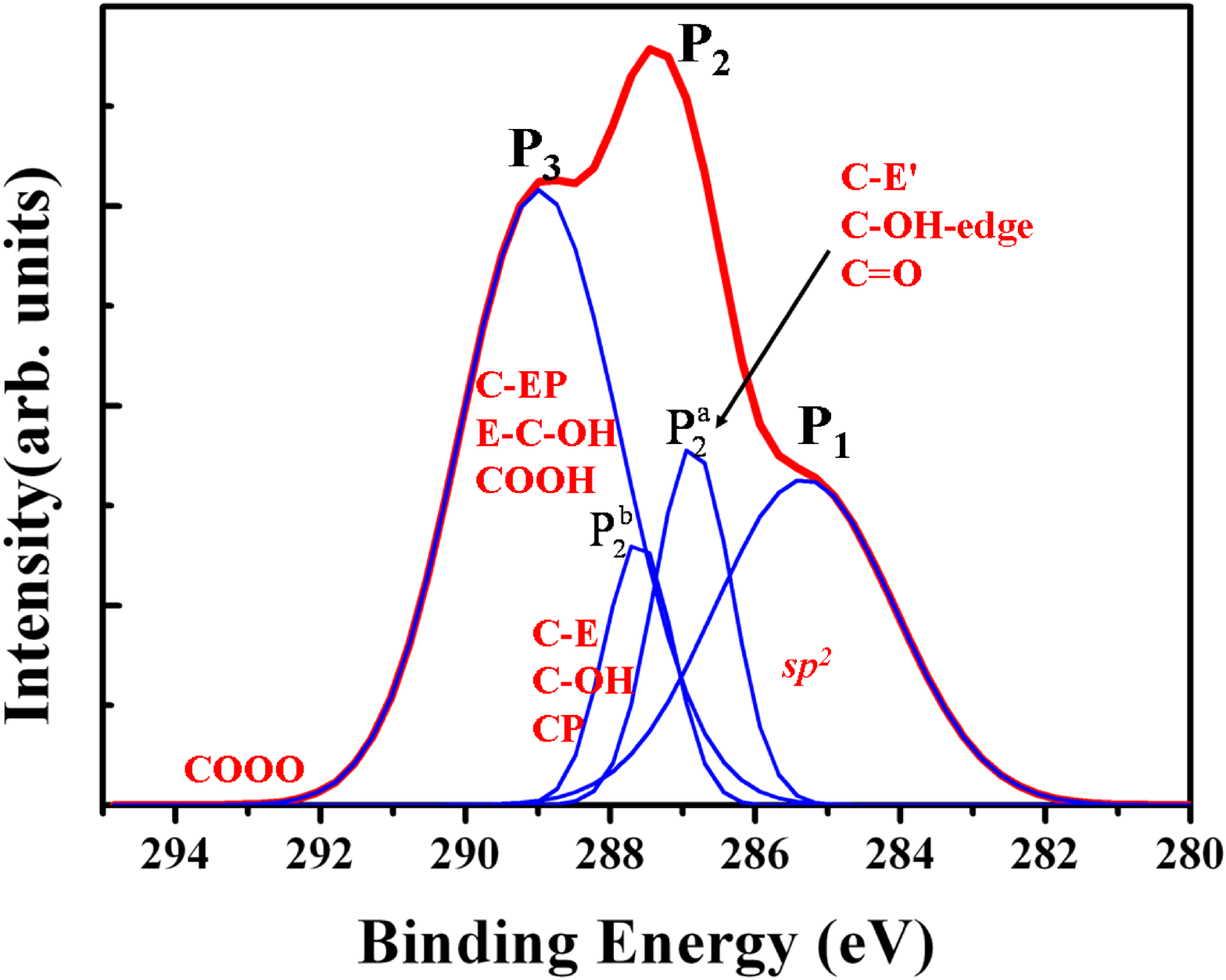}
\caption{Schematic experimental XPS curve and our first principles
assignment.} \label{fig:GO-XPS}
\end{figure}

In some experimental XPS data, there is a small peak with a R-CCS
more than 3 eV.\cite{Dekany:06:CM-XPS} Our calculations suggest that
it can be related to carbon atom connected to three oxygen, possibly
at GO edges. \cite{EPAPS} Until now,
as shown in Figure \ref{fig:GO-XPS}, we reach a unified assignment
for experimental XPS. The lowest peak $P_1$ comes from $sp^2$ carbon.
Peak $P_2$ has two contributions: the lower one comes from edge groups
and edge-like group C-E$^\prime$, and the higher one is contributed by
epoxy and hydroxy groups. Peak $P_3$ mainly comes from C-EP and E-C-OH,
with also small contributions from edge COOH groups.

With the proper first-principles assignments, the XPS intensity
evolution observed in experiments \cite{Dekany:06:CM-XPS,
Lee:06:JACS} can also be well understood. First of all, when the
degree of oxidation increases, the peak of $sp^2$ carbon ($P_1$)
should decrease or even disappear for completely oxidized samples.
At the same time, an increase of $P_3$ compared to $P_2$
was observed experimentally, \cite{Dekany:06:CM-XPS,Lee:06:JACS} which is because
more EP and E-C-OH groups appear during the oxidation. Jeong et
al. \cite{Lee:06:JACS} observed an increase of $P_2^a$ and a
decrease of P$_2^b$ during oxidation, which can also be understood
with our assignments. In fact, during the increase of the degree
of oxidation, the possibility for epoxy groups to align in a line
increases, which leads to more C-E$^\prime$ groups ($P_2^a$).
When GO is heated, a significant decrease of $P_3$ has been observed.
\cite{Lee:06:JACS} The dissociation of EP to CP should have made an
important contribution to it.

\section{Conclusions}

Our DFT calculations provide relative C 1$s$ core
chemical shifts for possible oxidation species on GO, whose binding
energy follows the order: C \emph{$sp^2$} $<$ edge groups (such as
C=O, C-OH-edge, and C-E$^\prime$) $<$ C-OH and C-E $<$ C-EP, COOH, and
E-C-OH $<$ COOO. These results lead to a reassignment of
experimental XPS spectra: (1) $P_2$ is mainly contributed by C-E and C-OH, and
they are not distinguishable; (2) edge groups and edge-like C-E$^\prime$
groups has lower binding energies compared to other $P_2$ groups;
(3) although it has been previously used to explain $P_3$ in XPS experiments, C=O actually has a
low binding energy corresponding to $P_2$;
(4) $P_3$ mainly comes from new species C-EP and E-C-OH; (5) even higher
binding energy may obtained from COOO.

Based on our first principles spectroscopic
understanding, we reach the following two important conclusions about
the atomic structure of GO: (1) epoxy and hydroxy groups
can be closely packed together forming epoxy pairs and epoxy-hydroxy pairs in
highly oxidized samples; (2) as the main species contributed to the lower energy part of $P_2$,
aligned epoxy group (C-E$^\prime$) represents an important GO structure motif,
which should be considered in all GO property studies and not only in oxidation induced cutting.
Theoretical simulations also lead to an understanding of the experimental evolution of XPS for
different samples, which provides us some insights to the oxidation processes
of GO. Our results demonstrate that first principles computational spectroscopy
is a very promising tool in nano structure research.

\begin{acknowledgements}
This work is partially supported by NFSC (20803071, 20933006, 50721091),
by MOE (FANEDD-2007B23, NCET-08-0521),
by CAS (KJCX2-YW-W22), by the National Key Basic
Research Program (2006CB922004), and by USTC-SCC, SCCAS, and Shanghai Supercomputer Center.
\end{acknowledgements}

\end{document}